\documentclass[twocolumn,widetext,preprintnumbers,amsmath,amssymb]{revtex4}
\usepackage{graphicx}
\usepackage{dcolumn}
\usepackage{bm}

\begin{document}

\title{Formal vs self-organised knowledge systems: a network approach}
\author{A. P. Masucci}
\altaffiliation{Instituto de F\'{\i}sica Interdisciplinar y Sistemas Complejos IFISC (CSIC-UIB), E-07122 Palma de Mallorca,Spain}

\begin{abstract}
In this work we consider the topological analysis of symbolic formal systems in the framework of network theory. In particular we analyse the network extracted by \emph{Principia Mathematica} of B. Russell and A.N. Whitehead, where the vertices are the statements  and  two statements are connected with a directed link if one statement is used to demonstrate the other one. We  compare the obtained network with other directed acyclic graphs, such as a scientific citation network and a stochastic model. We also introduce a novel topological ordering for directed acyclic graphs and we discuss its properties in respect to the classical one. The main result is the observation that formal systems of knowledge  topologically behave similarly to self-organised systems.
\end{abstract}
\maketitle

\date{\today}

\section{\label{sec:level1}Introduction.\protect}

A formal system is a set of a formal language and inferring rules that allows to create statements. Mathematics and logics are important examples of formal systems. In particular mathematics is an axiomatic formal system  that  develops theorems and propositions starting from a set of axioms. The importance of formal systems in nature resides on the fact that many scientific disciplines are developed as or are based on a symbolic formal system. Also many philosophic theories are expressed as formal systems, that is a set of axioms/definitions and propositions/corollaries/theorems. In this sense studying the properties of formal systems is equivalent to understand an important part of the  knowledge machine structure \cite{skirms}.

Formal systems can be described at a structural level within the framework of network theory. If we consider the statements of a formal system, definitions, axioms, theorems, etc.,   as  vertices and we connect by a directed link two vertices if one statement is used to demonstrate another one (for instance a definition is used to demonstrate a theorem), we obtain a so called \emph{directed acyclic graph} (DAG hereafter) \cite{dag}.

Formal systems are only  some of the main  references for knowledge production. Other knowledge systems in nature are far less formal, but still very important. Recently it has been  shown that an encyclopedia  is a self-organised system that grows and evolves as a biological system \cite{ss}. Moreover we can consider as a growing knowledge system the ensemble of scientific publications. In particular, as we explain better below, the directed network of scientific  citations  is keen to be structurally compared to  formal systems.

A DAG is a directed graph without closed cycles. DAGs are widely spread in nature. As an example each tree is a DAG (the viceversa doesn't hold). Moreover in nature we find  that transportation networks as rivers, cardiovascular and respiratory networks, plant vascular, food webs and root systems are DAGs, within knowledge systems  citation networks of papers or patents, legal cases are DAGs  and between artificial networks we find that electric circuits, feed-forward neural and transmission networks are also DAGs \cite{bn,kn,so}.

In some of these networks, such as the scientific and patent citations, or the legal cases, the acyclicity arises from the temporal ordering of the vertices. For instance papers can cite just older papers, as legal cases can cite just older legal cases \cite{legal}, hence closed cycles are difficult to form. In other networks, such as the formal systems,  the acyclicity of the graph arises from causality issues, so that  proposition A cannot be used to demonstrate  proposition B that is used to demonstrate proposition A.

While other DAGs in nature have already received the attention of the scientific community, a statistical analysis of a formal system of knowledge is still missing. One of the possible reasons for such a lack is that often mathematics books are quite small, with just a few hundreds  statements, and don't allow a statistical analysis.

In this research we present the topological analysis of the first volume of Principia Mathematica of B. Russell and A.N. Whitehead \cite{pm}, possibly the largest formal system in nature. Just in the first volume there are 2125 statements with 6805 connections.  We show that the topology of the network is non trivial. In particular the network is scale-free in the out-degree distribution and exponential in the in-degree distribution and reveals \emph{small world} properties \cite{dw}. We show that these are common  features with the citation networks.
To underline the similarity and differences between the formal system web and the citation network we run a parallel analysis of the citation network extracted by the journal ``Scientometric",  that has a similar number of vertices and links. We compare all the results with the DAGs obtained by shuffling the network within a configuration model \cite{boguna} and with a simple stochastic model that reproduces the main statistical properties of the DAGs in consideration.

\subsection{Ordering}

In DAGs where the nodes have an explicit time dependence, as in the citation graphs, an ordering for the graph vertices is straightforward just considering the vertices in a sequence where older vertices come before younger ones \cite{kn}. Nevertheless, even without considering the vertices time dependency,  DAGs  vertex ordering  is naturally induced by the topology of the network itself.

\begin{figure}[!ht]\center
                \includegraphics[width=0.4\textwidth]{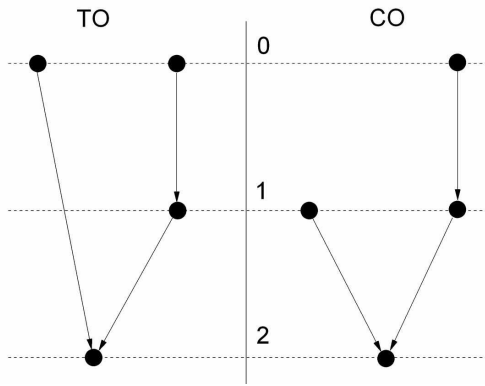}
 \caption{\label{f1} Difference between the topological ordering TO and the causal ordering CO for a simple DAG.}
 \end{figure}

\begin{figure*}[!ht]\center
                \includegraphics[width=0.95\textwidth]{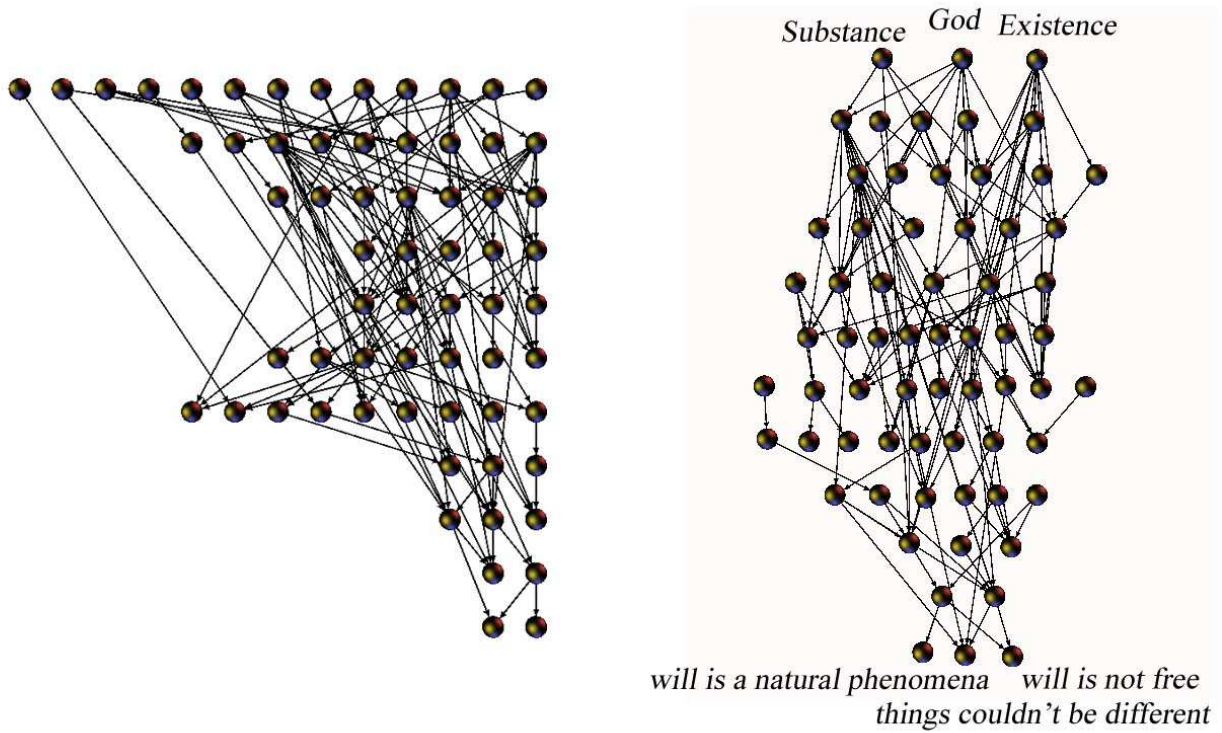}
 \caption{\label{f2} The DAG extracted from the first volume of the Ethics of Spinoza. On the left the TO of the network and on the right the CO of the network. We notice how in the CO there are 12 levels while in the TO just 11, this reducing the degeneracy of the levels in the CO representation. For the CO representation we show the definition names at the top level and the  proposition names at the bottom level. }
 \end{figure*}

 \emph{Topological ordering} (TO hereafter) is a fast algorithm (the time is linear with the number of edges) to sort the vertices of a DAG, so that for each directed edge the vertex at the beginning of the edge comes before the vertex at the end of the edge \cite{to}  (see Fig.\ref{f1} and Fig.\ref{f2}, left panels). The TO induces a layering in the network so that each vertex acquires an additional degree of freedom, given by the level the vertex belongs to. The TO induces a degeneracy in the levels, in the sense that many vertices can belong to the same level. For instance all vertices with indegree 0 belong to level 0 after a TO.

If TO makes sense in the contest of task management where it was introduced, it's not easy to interpret it in physics. Considering the kind of problems we are dealing with, for instance formal systems, vertices with indegree 0 are definitions or axioms. Then the TO would assign all definitions at level 0. In our opinion it would be better a sorting where the definition is positioned at the level before it is first used in a demonstration. Then we call  \emph{causal ordering} (CO hereafter) an ordering where each node is positioned in the level before the node it is connected to, with the constraint that all links have the same causal direction. In our knowledge CO has not been introduced before. To have a better idea of the difference between TO and CO  we show in Fig.\ref{f1} a simple DAG ordered via TO in the left panel, and via CO in the right panel.

There are more differences between TO and CO  than it could appear at a first glimpse with a simple graph. First of all if the vertex level for TO is well defined, for CO it can vary at each realization of the ordering. Hence in the case of CO we  talk about the average level $<l>$ of a vertex over a sufficient number of realization of the CO. Then the average standard deviation $<\sigma_l>$ of the levels divided by the average maximum level of the DAG, $l_{Max}$, quantifies the stability of the CO and is a measure of the complexity of the DAG itself. This number must be small in order for the CO to be meaningful. Nevertheless we find some good reasons to propose the CO as an interesting way to sort the DAG vertices. First of all the maximum level of the DAG for CO is larger or equal to the maximum level of the DAG for TO and this reduces the degeneracy of the levels. Moreover since the CO has a  physics meaning, the resulting sorted network gives at a glimpse interesting information about the network itself. To understand this point we show in Fig.\ref{f2} a DAG representation of the first volume of the Ethics of Spinoza \cite{ethics}, showing in the left panel the TO and in the right panel the CO of the network. At an eye inspection we can catch the higher definitions at a hierarchical level and the deepest conclusions of the book.

\subsection{Notation}

A directed network  \emph{\textbf{$G$}} is a couple $\{V,E\}$ of $V$ vertices connected by $E$ directed links. A network is completely specified by its adjacency matrix $A=\{a_{i,j}\}$, $i,j=1,...,V$, whose elements are $a_{i,j}=1$ if there is a directed link from vertex $i$ to vertex $j$, $a_{i,j}=0$ otherwise. The first order correlations in a network are specified by its degree or connectivity distribution $P(k)$. The degree or connectivity of a vertex $k$ is defined as the number of its first neighbours. Then the out-degree of vertex $i$ is defined as $k_i^{out}\equiv\sum_{j}^Va_{i,j}$ and the in-degree of vertex $i$ is $k_i^{in}\equiv\sum_j^Va_{j,i}$.

\section{Empirical analysis}

In our study we compare the analysis of three networks: the network extracted from the first volume of Principia Mathematica, a citation network and a stochastic model. Also we compare the analysis of the real network with measures obtained by shuffling the network, while preserving the out and in-degree sequence.

\subsection{Principia Mathematica}

In the first half  of the last century a major debate took place in the philosophic and scientific community on the possibility that mathematics could be represented by a unique set of axioms. To demonstrate that, B. Russell and A.N. Whitehead began the cyclopic mission of writing an axiomatic symbolic system describing the fundamental mathematic truths. A few years after the revised second edition of the three volumes of Principia Mathematica, G\"odel formally demonstrated that axiomatic systems are intrinsically incomplete putting a point to a long discussion \cite{go}.
Nevertheless in this way Russell and Whitehead left to the world an opera that is the largest symbolic formal system ever written.

In this work we analyse the first volume of the opera, in its first edition of 1910. The extraction of the network from the PDF version \cite{pms} takes advantage from the fact that the number each statement is named with is preceded by the symbol ``$*$".

The network is composed by 2125 vertices and 6805 directed edges. The arrow of the edges follow a causal direction so that, for instance, the link goes from the definition to the proposition. In this representation the network has 285 \emph{roots}, links with in-degree 0, that are axioms and definitions.

\begin{figure*}[!ht]\center
                \includegraphics[width=0.6\textwidth]{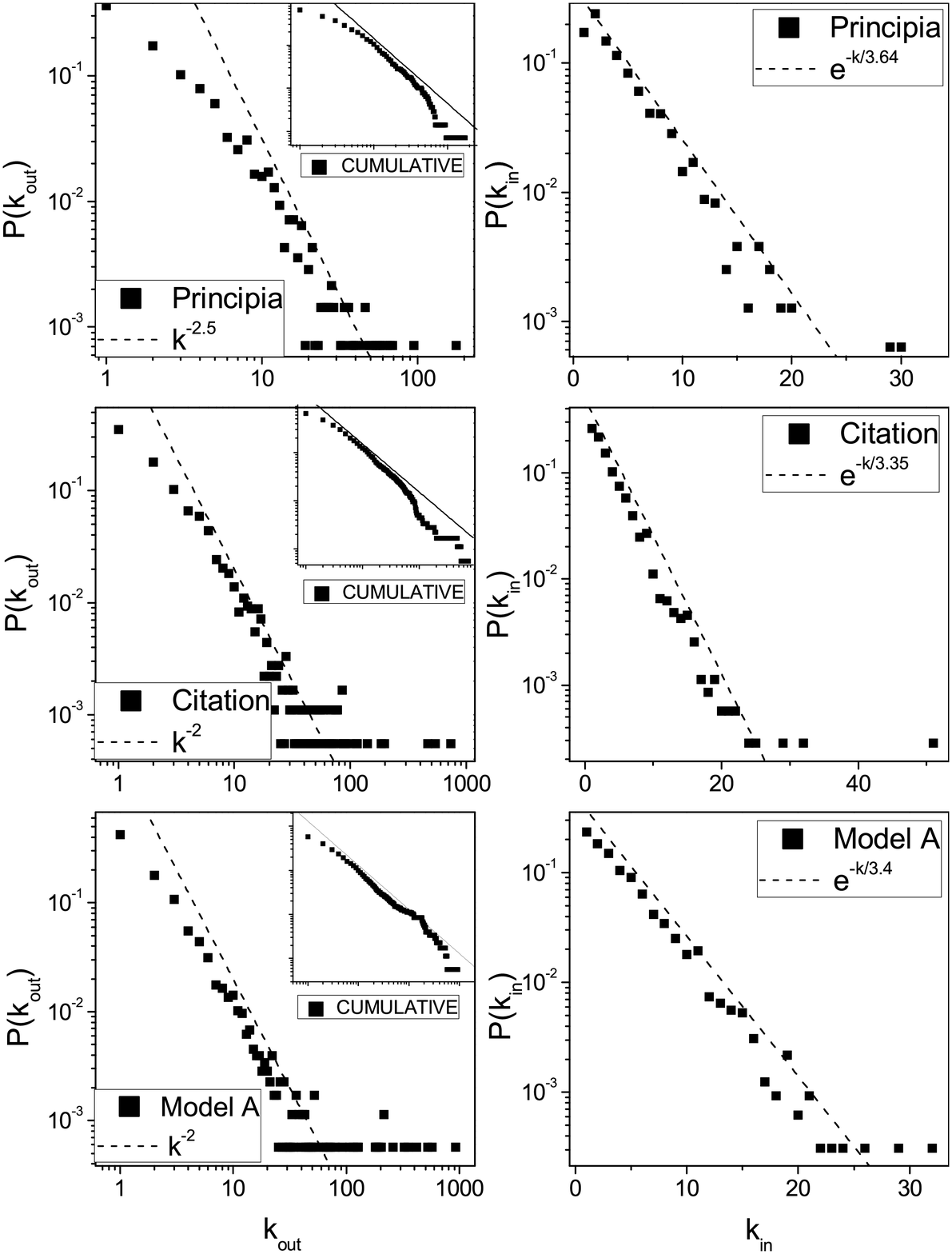}
 \caption{\label{f3} Out and in-degree distributions for the three networks in consideration. In the insets the relative cumulative distributions are displayed. Top panel: the Principia Mathematica. Middle panel: the citation network. Bottom panel: model A.}
 \end{figure*}

In the top-panel of Fig.\ref{f3} we show the out and in-degree distribution of the network. We can see that the network is scale-free in the out degree distribution for many decades with an exponent around -2.5, $P(k_{out})\propto k_{out}^{-2.5}$.  On the other hand the in-degree distribution is well fitted by an exponential function $P(k_{in})\propto \exp(-k_{in}/\langle k_{in}\rangle)$, where $\langle k_{in}\rangle=\sum_ik_i^{in}/V$ is the average in-degree. Those results are strikingly similar to the classical results reported about citation networks \cite{5,vas}.

The out-degree distribution for citation networks has been explained on the basis of multiplicative models mimicking a popularity phenomena, i.e. \emph{rich get richer} \cite{5}. It looks like that a similar reasoning can be applied to the case of formal system networks where a hierarchy of statements emerges and where hubs form. Those hubs reflect the fact that some of the statements are widely used through the whole system, i.e. they have a global influence on the structure of the network, while many statements are used just to develop local structures. Moreover the diameter of the network is 11 that is close the logarithm of the network size, hence revealing \emph{small world} properties of the network \cite{dw}.

\subsection{Citation Network}
The statistics of citations in scientific journals has long been considered by the scientific community \cite{price,redner}. Network theory appears to be an    appropriate framework where to study this kind of phenomenology, since it naturally includes the statistics of citing papers to the one of cited papers. In particular it has been shown that, considering the link arrows going from citing papers to cited papers, the out-degree distribution for citation networks is compatible with a scale-free distribution, while the in-degree distribution can be fitted by an exponential distribution \cite{vas}.

It is interesting to notice that even if the DAG structure of a citation network can emerge by the time sequence of the papers, it can also be driven by causality reasons. In fact citation networks are knowledge structure that have many similarities with the formal systems considered in this work. The reason is that each scientific paper doesn't start from scratch to demonstrate something, but from an already present structures of papers that are considered reliable since they are already published. Then the parallel between formal statement/paper is straightforward, if you use paper A to demonstrate paper B that was used to demonstrate paper A, you probably fall in a logical contradiction. The main difference between formal and citation networks is that the knowledge system of papers is a collective phenomena, while the formal ones often come from a single writer that recollects and formalise/crystalise the knowledge of the time.

In this work we consider the network of citation extracted from the journal \emph{Scientometrics}, from 1978 till 2006 \cite{cn}. The choice of this network between the many of them that are publicly available, comes from the observation that this network has a size (number of vertices and links) of the same order of the one of Principia Mathematica. In particular it has 3772 vertices and 12719 directed links. In the middle panel of Fig.\ref{f3} we show its out and in-degree distribution. As we can see the out-degree distribution is compatible with a scale-free distribution over a large range of the degree with exponent close to -2, $P(k_{out})\propto k_{out}^{-2}$. On the other hand the in-degree distribution is exponential and it is finely fitted by the function $P(k_{in})\propto \exp(-k_{in}/\langle k_{in}\rangle)$.

\subsection{A stochastic model}\label{sm}

To complete our analysis we introduce a growing stochastic model, that we call \emph{model A}, based on Price's cumulative advantage model (CAM hereafter) \cite{price2}, that mimics the main topological properties of the considered DAGs. The CAM is pretty similar to the preferential attachment model \cite{5}, with the difference that, being the network  directed,  the attachment probability is proportional to the out-degree plus one instead of just the out-degree. To create a multi-rooted DAG we insert uniformly nodes with in-degree 0 in the growth process of the CAM.

To create a network with $N$ vertices, the model goes like that. We start with $N_0$ disconnected vertices. Then at each time step we generate a new vertex and we draw $m$ directed links from already existing vertices to the new one.  Vertex $i$ is selected for attachment accordingly to the probability $\Pi_i=\frac{k^{out}_i+1}{\sum_j (k^{out}_j+1)}$. During the whole process, uniformly in time, we generate $M_0\ll N$ vertices with in-degree 0 and we attach them to the new vertex. The number $m$ is extracted from a distribution that mimics the in-degree distribution of the real systems in Fig.\ref{f3}: $P(m)\propto e^{-\frac{m}{3.4}}$.

In the bottom panel of Fig.\ref{f3} we show that this model produces a network that is topologically very similar to the citation network and to the formal system network and then it can be used as a null model to analyse further the correlations within the real systems.

\subsection{Configuration model}\label{cm}

Some models have already been proposed as shuffling procedures for DAGs \cite{so,kn}. Here we consider the following one: we choose two links of the network $(i,j)$ and $(h,k)$ and we consider the network after the rewiring process that change the two links with the links $(i,k)$ and  $(h,j)$, with the bias that multiple links are not allowed. If the resulting network is still a DAG we keep the change and go on with the same procedure for a number of times of the order of the number of links. This shuffling model keeps the out and in-degree sequence of the network unchanged, so that the resulting network keeps the same properties of the original networks until the first order correlations, and it randomises higher order correlations.

\subsection{Comparison}

\begin{figure}[!ht]\center
                \includegraphics[width=0.4\textwidth]{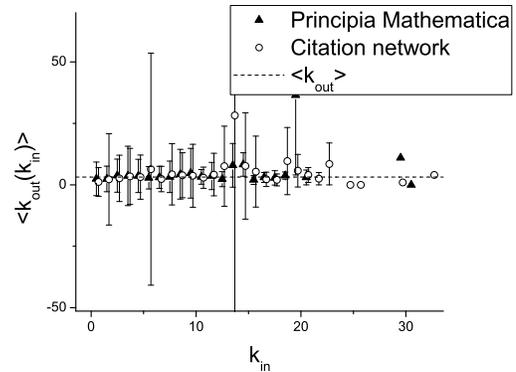}
 \caption{\label{f4} Correlations between out-degree and in-degree for the vertices of the Principia Mathematica and the citation network. The dashed line represent the average out degree for the networks. The average out-degrees are close and is not possible to resolve the two lines at this scale.}
 \end{figure}

The main topological properties of the considered networks we introduce in the last subsections can be   contextualised in a rich get richer dynamics framework. However  the linear preferential attachment has been explicitly measured for citation networks \cite{redner}. To go further in the analysis, in \cite{kn} the correlations between out and in-degree for the vertices have been analytically computed for the configuration model and it has been found that they don't differ in a significative way from the correlations of real citation networks. In fact we find that also in our real networks the out-degree and the in-degree of the vertices are not correlated as we show in Fig.\ref{f4}.

To characterise the complexity of DAGs we find more interesting the approach in \cite{bn}, where the tree-like structure of DAGs is considered. In particular the complexity of DAGs can be characterised in terms of the balance of the network subtrees.

Tree shape can vary between  completely balanced, i.e. a symmetric tree, to completely unbalanced, i.e. a comb-like tree (for instance see \cite{alex} for a visual reference). In general a complex tree shape is in between those two extremes. It has been shown that the functional relation between the subtree size and its cumulative measure is useful to quantify the tree balance. In particular, given a DAG, we define the subtree size $A_i$ of vertex $i$, the size of the subset of vertices $\{S_i\}$, that are the vertices that can be reached from vertex $i$ through directed links, including vertex $i$ itself. Then the cumulative subtree size is defined as $C_i\equiv \sum_{S_i}A_i$. As we show below, $A_i$ embeds important information about higher order correlations in the DAG.

\begin{figure*}[!ht]\center
                \includegraphics[width=0.6\textwidth]{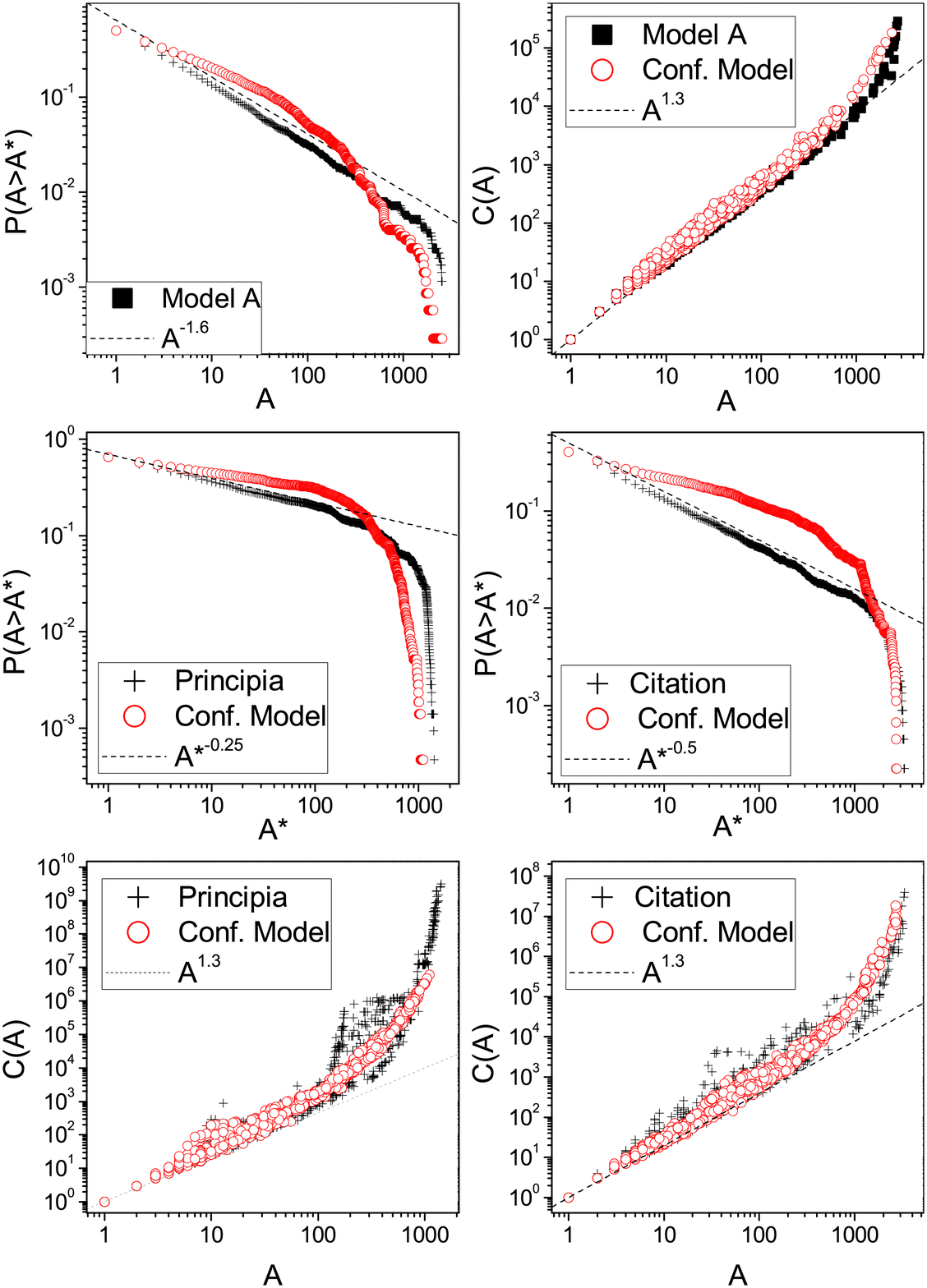}
 \caption{\label{f5} Top panels: on the left the cumulative subtrees size distribution $P(A>A^*)$ for \emph{model A} and its configuration model; on the right panel the functional relation between the subtrees size and the cumulative subtrees size $C(A)$ for \emph{model A} and its configuration model.
 Middle panels: the cumulative subtrees size distribution $P(A>A^*)$ for the Principia Mathematica network and its configuration model (left panel) and for the citation network and its configuration model (right panel). Bottom panels: functional relation between the subtrees size and the cumulative subtrees size $C(A)$ for the Principia Mathematica network and its configuration model (left panel) and for the citation network and its configuration model (right panel).}
 \end{figure*}

We start with the analysis of the stochastic model, since it shows robust statistical behaviours. In the top panels of Fig.\ref{f5} we show the subtrees size cumulative distribution $P(A>A^*)$ and the functional relation between the subtrees size and its cumulative function $C(A)$, for \emph{model A} and the network obtained with the configuration model outlined in Subsec.\ref{cm}. For \emph{model A}, $P(A)$ displays a robust scale free behaviour, $P(A>A^*)\propto {A^*}^{-(\gamma-1)}$, until very large scales, with exponent $\gamma\approx 1.6$. $C(A)$ also is scale-free until very large scales, $C(A)\propto A^{1.3}$. This kind of behaviour was already noticed in \cite{alex,bn} for phylogenetic trees and river networks. It is interesting to notice that in the configuration model the scale-free behaviour for the subtrees size distribution  is not preserved, while $C(A)$, representing the balance of the DAG subtrees, preserves the same power law with small deviations for large scales. Since the configuration model preserves the out and in-degree sequence of the vertices, we can deduce that the scale-free behaviour of $P(A)$ is associated to higher order correlations, possibly associated to  the preferential attachment mechanism in model A.

In the middle panels of Fig\ref{f5} we show the cumulative degree distribution for the real networks in consideration and for the relative shuffled networks. In the left panel we show the cumulative distribution for the subtrees size for the Principia Mathematica network. It is scale-free until moderately large scales, $P(A>A^*)\propto {A^*}^{-(\gamma-1)}$, with $\gamma=1.25$, and then it decays exponentially for larger scales. In the same plot we show the same quantity measured in the configuration model. We can appreciate that the scale-free behaviour is partially destroyed in the shuffled net. In the right panel we show the cumulative distribution for the subtrees size for the citation network. It is scale-free until  large scales, $P(A>A^*)\propto {A^*}^{-(\gamma-1)}$, with $\gamma=1.5$. In this case the scale-free distribution is completely destroyed after the shuffling process of the configuration model.

In the bottom panels of Fig.\ref{f5} we show the relation between the subtrees size and the cumulative subtrees size for the vertices of the real networks in comparison with the same quantity measured for the shuffled networks via configuration model. Until moderately large scales, $C(A)$ follows a power law with  exponent 1.3, similar to the one of the stochastic model and then depart from the power law for larger scales. It is interesting to notice how $C(A)$ differ between real and shuffled networks for a more structured behaviour, characterising the complexity of the local structures for the real networks in comparison to the shuffled ones, but it preserves the main functional behaviour.

\section{Discussion and conclusions}

In this work we presented the first statistical analysis of a formal system, that is the network extracted from the Principia Mathematica. Through the parallel comparison with two other DAGs, a citation network and a growing stochastic model based on preferential attachment, we could show some peculiar traits of the networks in consideration. In particular we showed how an analysis focused on the complexity of the subtrees of a DAG can give more information than an analysis based on out-degree-in-degree correlations.

Interestingly enough we found that the two real networks considered in this research have substantial common traits.
At a global level they're characterised by a scale-free distribution for the out-degree and exponential distribution for the in-degree. This allows us to classify them both in a network family that can be described by a class of models such as the CAM (Subsec.\ref{sm}), that is the precursor of the Barabasi-Albert model \cite{5} and that shares many common traits with it.

With the help of model A and of the configuration model, we notice that the scale-free subtrees size distribution $P(A)$ is not directly related to the scale-free degree distribution, but  it can give information about second or higher order correlations and it possibly relates to the preferential attachment mechanism acting during the growth of the network. This can be an important observation. In fact, as already noticed, linear preferential attachment has been directly measured in citation networks \cite{redner}, but it is improbable that a formal system will be found to be large enough to allow such a measurement. Then an undirect proof like the one we propose can be an important hint for the understanding of the growth of formal systems.

A statistical approach to formal systems is not only important for speculative reasons, i.e. as the structural description of a natural phenomena, but it is useful for information retrieval, that is to have a clear vision of the hierarchy and the complexity embedded in the system. As an example we can think about the still open and fascinating problem of the 5th postulate of Euclid \cite{euclid} and how the Euclid's Elements analysis  as a DAG could help to better understand its structural role. In this sense we introduced a novel algorithm for the ordering of DAGs, the CO, and we showed that it better catches the complexity of the system, reducing the degeneracy levels   in the ordering in respect to the classical TO and giving interesting information about the formal system just at an eye inspection.

\begin{acknowledgments}
Supported by Ministerio de Ciencia e Innovaci\'on and Fondo Europeo de Desarrollo Regional through project FISICOS (FIS2007–60327).
I would like to thank D. Villone, E. Hern\'andez-Garc\'ia and V. M. Egu\'iluz for the interesting conversations on the topic and A. Mura to be inspirational about the analysis of the Ethics of Spinoza. The Gephi freeware software was used to produce Fig.\ref{f2} \cite{gephi} and the software Inftyreader was used to digitalise the PDF version of the Principia Mathematica \cite{infty}.

\end{acknowledgments}


\end{document}